\documentclass{iopart}
\begin{document}

\def\beq{\begin{eqnarray}}
\def\eeq{\end{eqnarray}}
\def\defn{\buildrel \rm def \over =}
\def\pre{\textit{Preprint~}}

\def\s{\mbox{\boldmath$\displaystyle\mathbf{\sigma}$}}
\def\bp{\mbox{\boldmath$\displaystyle\mathbf{\pi}$}}
\def\be{\mbox{\boldmath$\displaystyle\mathbf{\eta}$}}
\def\J{\mbox{\boldmath$\displaystyle\mathbf{J}$}}
\def\K{\mbox{\boldmath$\displaystyle\mathbf{K}$}}
\def\P{\mbox{\boldmath$\displaystyle\mathbf{P}$}}
\def\p{\mbox{\boldmath$\displaystyle\mathbf{p}$}}
\def\hp{\mbox{\boldmath$\displaystyle\mathbf{\widehat{\p}}$}}
\def\x{\mbox{\boldmath$\displaystyle\mathbf{x}$}}
\def\X{\mbox{\boldmath$\displaystyle\mathbf{X}$}}
\def\0{\mbox{\boldmath$\displaystyle\mathbf{0}$}}
\def\bv{\mbox{\boldmath$\displaystyle\mathbf{\varphi}$}}
\def\bx{\mbox{\boldmath$\displaystyle\mathbf{\xi}$}}
\def\bs{\mbox{\boldmath$\displaystyle\mathbf{\sigma}$}}
\def\bc{\mbox{\boldmath$\displaystyle\mathbf{\chi}$}}

\def\hbv{\mbox{\boldmath$\displaystyle\mathbf{\widehat\varphi}$}}
\def\hbxi{\mbox{\boldmath$\displaystyle\mathbf{\widehat\xi}$}}
\def\bn{\mbox{\boldmath$\displaystyle\mathbf{\nabla}$}}
\def\bl{\mbox{\boldmath$\displaystyle\mathbf{\lambda}$}}
\def\br{\mbox{\boldmath$\displaystyle\mathbf{\rho}$}}
\def\1{1}
\def\ar{\stackrel{\hspace{0.04truecm}grav. }{\mbox{$\longrightarrow$}}}

\title[A freely falling frame at the interface of
gravitational and quantum realms]
{A freely falling frame at the interface of
gravitational and quantum realms}

\author{D V Ahluwalia-Khalilova}  

\address{
 	Ashram for the Studies of the Glass Bead Game (ASGBG)\\
 	Ap. Postal C-600, Zacatecas, Zac 98060, Mexico\\
 	and\\
 	Center for Mathematical, Physical, and Biological Structure\\
 	of Universe (CIU), Ap. Postal C-612, Department of Mathematics\\
 	University of Zacatecas (UAZ), Zacatecas, Zac 98060, Mexico}

\eads{d.v.ahluwalia-khalilova@heritage.reduaz.mx }

\begin{abstract}\\
I briefly argue for logical necessity 
to incorporate, besides $c$, $\hbar$, 
two fundamental length scales in the symmetries associated with 
the interface of gravitational and quantum realms. 
Next, in order to clear the proverbial bush,
I discuss the CPT and indistinguishability issue related
to recent non-linear deformations of special relativity
and suggest why algebraically well-defined extensions
of special relativity do not require non-linear deformations. 
That done,  I suggest why the stable  Snyder-Yang-Mendes  
Lie algebra should be considered as a serious candidate for the symmetries
underlying freely falling frames at the interface of 
gravitational and quantum realms; thus
echoing, and complementing, arguments recently put forward by
Chryssomalakos and Okon.  In the process I obtain concrete form
of uncertainty relations which involve above-indicated length scales
and a new dimensionless constant. I draw attention to the fact
that because 
superconducting quantum interference devices can carry
roughly $10^{23}$ Cooper pairs in a single quantum state,  
Planck-mass quantum systems already exist in the laboratory. These may be
used for possible exploration of the interface of the
gravitational and quantum realms.
\\{~}\\
\textbf{Journal Ref.:} Class. Quantum Grav. {\bf 22} (2005) 1433-1450

\end{abstract}


\maketitle


\section{Introduction}

When one sets out to think about quantum gravity,
her/his first question perhaps ought to be: What meanings do 
`quantum' and 
`gravity' carry in any such theory?  
The fundamental significance of this question arises from the 
facts that \cite{Dirac:1958pam,Wigner:1962bww,Weinberg:1995volI}: 
(a) The Heisenberg's 
fundamental commutators $[x,\,p_x] =
i \hbar$, $\ldots$ lie at the heart of wave-particle duality and affect
the entire
quantum mechanical framework of fields and particles; and that (b) Poincar\'e
spacetime symmetries \textemdash~
as an algebraic representation of constancy of speed of light $c$ for
all inertial observers \textemdash~
 not only define the notion
of particles, 
but also suggest the equality of inertial and 
gravitational masses  \cite{Weinberg:1964ew}.
Furthermore \cite{Doplicher:1994zv,Ahluwalia:1993dd}, 
Doplicher \textit{et al} and the author have independently argued that
interplay of the uncertainty relations, 
$\Delta x\, \Delta p_x \ge \hbar/2$, $\ldots$ (following from the 
fundamental commutator) and Einsteinian gravity,
renders spacetime measurements non-commutative. These observations
already hint that meanings of quantum and gravity may  undergo
conceptual and mathematical modifications at the interface of
gravitational and quantum realms (\texttt{IGQR}).

A precise and concrete 
answer to the opening question comes from the stability analysis 
of the associated algebraic structures \cite{VilelaMendes:1994zg} \textemdash~
where it may be noted that from a  physicist's point of view 
a Lie algebra is considered stable
if infinitesimal perturbations in its structure
constants leads to isomorphic algebras (see, e.g,
\cite{Chryssomalakos:2001nd}). 
The analysis presented in    \cite{VilelaMendes:1994zg}, and
now confirmed and extended in  \cite{Chryssomalakos:2004gk},
says in essence, that
neither the Heisenberg algebra, nor the Poincar\'e algebra,
preserves its stability at the \texttt{IGQR}\footnote{To 
avoid confusion it is to be noted that the stability 
of the Poincar\'e algebra away from 
the gravitational realm refers to the kinematical group
of the tangent space to the spacetime manifold and not
to the group of motions in the manifold itself. 
A ``harmless'', i.e. devoid of physical implications,
instability also exists for the Heisenberg algebra.
It is an artifact of singling out $x$, as compared,
say, to $\exp(i x)$, as a physical observable. 
While we do not deal here with quantum deformations,
it is worth noting, as has been pointed out to us by one
of the referees, that a quantum deformation of a group
is a stabilization in the domain of Hopf algebras. From
this point of view also neither the Poincar\'e nor
the Heisenberg Lie algebra are stable. The reader is
referred to \cite{Celeghini:1990xx,Celeghini:1991bh,Celeghini:1990bf} 
for further details.
}. The stabilized Heisenberg-Poincar\'e algebra
asks for two additional length scales (to be discussed below)
in the same manner that the stability analysis of  \cite{VilelaMendes:1994zg} 
shows $1/c^2$ and
$\hbar$, without giving their numerical values, as parameters
required by the stabilization of the Galilean and classical  kinematics.
The stabilization, as reviewed in  
\cite{VilelaMendes:1994zg,Chryssomalakos:2004gk}, leads respectively
to special relativity (with $1/c^2$ as the deformation
parameter), and quantum mechanics (with $\hbar$ as the deformation
parameter). For earlier references which obtain similar results,
reader's attention is drawn to  \cite{Bayen:1977pr,Flato:1982yu}.
In addition, Faddeev has made the observation that general
relativity may be viewed to arise from special relativity with
Newtonian gravitational constant $G$ serving
as the deformation parameter \cite{Faddeev:1988LDX}. 
The paradigm of Lie-algebraic deformations to obtain stable theories 
claims not only historical success \textemdash~
in retrospect,
as having the power to have predicted relativistic and quantum revolutions
\textemdash~
but it is also the theory which identifies the fundamental constants
underlying their Lie algebraic structure.

There are two other possible answers to the questions asked.
The Lie-algebra deformations, leading to a modification
of the Heisenberg and Poincar\'e algebra,
 considered in the above-cited
works of Mendes  \cite{VilelaMendes:1994zg}  and that
of  Chryssomalakos and Okon \cite{Chryssomalakos:2004gk}
are in the classical sense of Nijenhuis and Richradson
\cite{Nijenhuis:1967arw}. These deformations, being minimal 
in the sense that they still preserve the Lie algebraic structure,
perhaps capture the essence of modifications to the notions
of quantum and gravity in \texttt{IGQR}.
It is possible that still 
higher order corrections/modifications occur in the 
context of q-deformations  or quantum groups \cite{Majid:2000qgt},
an observation already made by Mendes in the concluding 
paragraph of  \cite{VilelaMendes:1994zg}. These shall not be pursued
in this paper.

Another possible answer to the question asked is offered
by the phenomenological modifications of dispersion relations in
such a way that two, or more, additional deformation parameters
are introduced. These theories, considered
under a generic misnomer `doubly/triply special relativity',
despite significant 
amount of effort and publications, continue to suffer
from a lack of well-defined mathematical framework which assures
self-consistency. Additionally, they  have failed to provide 
a  satisfactory spacetime, or phase space, structure.  
For instance, Kowalski-Glikman and Smolin 
have compiled a list of four astrophysical and cosmological 
anomalies which seem to carry a single quantum-gravity origin
\cite{Kowalski-Glikman:2004kp}. In the same paper they 
suggest a new \textit{non-linear} deformation of the
Poincar\'e algebra
and put forward a `triply special relativity'.
Chryssomalakos and Okon \cite{Chryssomalakos:2004wc}
were immediately able to show that the proposed algebra
can be brought to a Lie \textemdash~
i.e. a \textit{linear} \textemdash~ form by an appropriate
identification of its generators, and that this linear
form was the same as that  arrived at by Mendes \cite{VilelaMendes:1994zg}
and Yang \cite{Yang:1947ud}. 
The  non-linearity in the Kowalski-Glikman and Smolin
proposal arises due to the implicit insistence that 
central charge[s] remain undeformed. Since there is no
physical or mathematical justification to make this
assumption, and since the motivation to introduce additional
invariant scales remains unaffected by this assumption, there
seems to be no reason to abandon linearity (i.e. Lie algebraic
framework). 
Therefore, on the positive side,
physical motivations provided
by literature on
`doubly/triply special relativity (DSR/TSR)' are indicative 
of a fundamental change required for notions of quantum and gravity
in \texttt{IGQR}.
On the discouraging side,
the DSR/TSR's theoretical framework remains far from a well-defined 
mathematical scheme and 
it carries dubious/incomplete interpretational elements.

Independently, as already noted above, 
it appears that any attempt which incorporates the
gravitational effects in quantum measurement of spacetime events 
leads to (a) a non-commutative spacetime and (b) to the 
associated modification of the fundamental commutators 
\cite{Ahluwalia:1993dd,Kempf:1994su,Ahluwalia:2000iw,Matschull:1997du}.
The latter, in the
framework considered in  \cite{Kempf:1994su,Ahluwalia:2000iw},
 leads to modification of the de Broglie wave-particle 
duality in such a manner that it saturates the matter wavelengths to
Planck length, $\ell_P \defn \sqrt{\hbar G/c^3}$. That is, irrespective of
the relative velocity of two inertial frames, $\ell_P$ does not Lorentz
contract. 
On the one hand this is a physical counterpart 
of Mendes' stability argument, and on the other
it immediately calls for modification of special 
relativity to incorporate not only the invariant $c$ but also
$\ell_P$. When one adds to this the Mendes' 
\cite{VilelaMendes:1994zg} stability argument for the 
combined Heisenberg-Poincar\'e algebra  
 one is forced to include yet another length scale.
That length scale may be tentatively identified with a 
large-scale cosmological property governed by 
$\ell_C = \sqrt{ 3 c^4/8 \pi G \rho_{vac}} \defn \sqrt{1/\Lambda}$, 
with $\Lambda$ being the cosmological constant.

The important thing for this paper is not that the length scales
take the values, $\ell_P$ and   $\ell_C$ but that there exist
two length scales: one in the extreme short-distance range, and the
other carrying astrophysical, or cosmological, scale.
In what follows this flexibility in the
identification shall be taken as implicit.
 
Freely falling frames being most appropriate realms
to establish the relativistic and quantum algebras, 
the primary aim of this paper  becomes to present a concrete 
modification to the notion of freely falling frames at the interface of
gravitational and quantum realms, and to address the related issues.
In the process we shall give \textit{algebraically} precise meaning to
the notions of `quantum' and `gravity' and point out some of the
most immediate implications.

In Section \ref{Sec:Proverbial}, as required by the above discussion,
I first attempt to clear the proverbial bush relevant to
address the issue of freely falling frames at the interface of the
gravitational and quantum realms. 
This effort also allows us to present a systematic methodology
to obtain wave equations. It applies 
not only for non-linear proposals, but
also to those 
Lie algebraic frameworks where the Lorentz sector remains intact.
Then, in
section \ref{Sec:FreeFall}, I return 
to the subject of as to what Lie algebraic structure  may 
the freely falling frame at the interface of gravitational 
and quantum realms carry.  This then provides a systematic 
step towards construction of a relativity for the \texttt{IGQR}
where spacetime acquires intrinsically quantum and gravitational 
character. That is, even in a freely falling frame there remain
intrinsically quantum and gravitational signatures. The spacetime
in IGQR requires not only $c$, but also $\hbar$, $\ell_P$, and
$\ell_C$ (and possibly a new dimensionless constant $\beta$
signifying a radical departure from some of the quantum relativistic
notions).
Section \ref{Sec:ConcludingRemarks} is devoted to concluding remarks.
With exception of section \ref{Sec:Proverbial}, where we set $\hbar$ and $c$
to be unity, we shall make them explicit in the remainder of this
paper.

\section{
CPT and indistinguishablity issue for non-linear deformations
of special relativity}  
\label{Sec:Proverbial}

In order to take the next logical step I find it necessary to 
first update and close an argument which I put forward 
a few years ago in  \cite{Ahluwalia-Khalilova:2002wf}.
Here, therefore, I summarize my views
on CPT and indistinguishablity issue 
as encountered in doubly
\cite{Amelino-Camelia:2000mn,Amelino-Camelia:2002vy,Magueijo:2001cr}, 
and now triply
\cite{Kowalski-Glikman:2004kp}, 
special relativity.
Before I address the matters of content, it may be useful
to make a few informal remarks and attend to  
a matter of nomenclature.
I take this liberty because, in my opinion, 
these remark set the essential tone of this paper
and because any future evolution of this
subject should be based on a more clear and well-defined
premise. 
On the indicated issue I do not give a categorical, 
or an unambiguous, answer (nor is one possible in a model-independent
manner); but,
instead, write this  paper in a manner which parallels 
the development of the ideas in the field and how, prematurely, one may
be tempted to claim results 
\cite{Kowalski-Glikman:2004kp,Amelino-Camelia:2002wr}
which on closer examination raise troubling 
questions\footnote{In
 the connection of \cite{Lukierski:2002df}, it is worth while to
note that the non-linearity of some commutators in quantum deformed 
algebra may be eliminated by means of an appropriate change of basis
but, then, one gets again the same Hopf algebra and not a Lie algebra.
For relevant details and construction of wave equations see 
\cite{Bonechi:1994ny,Bonechi:1994xyz}.} 
\cite{Chryssomalakos:2004wc,Grumiller:2003df,
Schutzhold:2003yp,Lukierski:2002df,
Rembielinski:2002ic,Toller:2003tz,Jafari:2003xt,Lukierski:2004jw}.

\subsection{Nomenclature} 

The special of `special relativity' refers to the circumstance that 
one restricts to a special class of inertial observers which move
with  relative uniform velocity. The general of `general relativity'  
lifts this restriction. The `special' of special relativity
has nothing to do with one versus two, or three,
invariant scales. It rather refers to the special class of inertial 
observers; a circumstance that remains unchanged in  special relativity
with two invariant scales. Taken to its [il]logical conclusion
it would mean that a theory of general relativity with
two invariant scales would be called `doubly 
general relativity'.
A detailed look at the 
algebraic structure of 
`doubly special relativity'
\cite{Amelino-Camelia:2002vy,Magueijo:2001cr}
 suggests that these are non-linear deformations of special relativity;
or, in the language of  \cite{Lukierski:2002df} 
a change of basis, associated with a nonlinear change of generators, 
in enveloping algebra.
The deformations are characterized with two 
invariant scales. 
The technically appropriate, though by no means a unique, 
nomenclature 
is thus: \textbf{n}on-linear deformations
of \textbf{s}pecial \textbf{r}elativity with \textbf{2} 
invariant scales, i.e., \texttt{NSR-2}. 
Many of the remarks I make here, though written in the context of 
\texttt{NSR-2}, remain valid for \texttt{NSR-3} as well.
Taking note of this observation, authors
of \cite{Chryssomalakos:2004wc} have expressed their opinion
in the following words, `we think the above term is conceptually 
inappropriate enough to warrant its abolishment ...'.
I agree.
The term they are referring to is `doubly/triply special
relativity'.

Yet, one may 
be tempted to preserve \texttt{DSR}, with \texttt{D} 
now meaning `deformed' rather than `double'.
 But, then, 
it does not distinguish between nonlinear and linear (i.e, Lie)
deformations. Nor does such an abbreviation extends naturally 
to triply special relativity.

\subsection{A mix of
deformations in algebra and transformation parameters}

In \texttt{NSR-2}s,
the underlying algebra for the rotations and boosts remains intact as the 
standard Lie algebra of the Lorentz group. The non-linear deformation
of the algebra is contained in the remaining sector.
As long as the motivation is to obtain a special
relativity with extended number 
of invariants  \textemdash~ 
from 1 to 2 
(or, even 3 as in    \cite{Kowalski-Glikman:2004kp}) \textemdash~
and as long as the Lie algebra framework provides  a well-defined framework
\cite{Chryssomalakos:2004wc}, the physical and mathematical justification
for invoking non-linear deformations seems to be too unwarranted 
a break from
the standard framework.\footnote{I concede that
to some extent this is a matter of taste
provided one has a well-defined 
mathematical and interpretational scheme.}
Furthermore, even though in \texttt{NSR-2}s the Lorentz algebra remains
intact, the boost parameter suffers a modification. This mix of
deformations in the underlying algebra and the associated 
transformation parameters complicates the theory significantly
enough that no fully satisfactory version of the theory exists beyond
the momentum space (and even there the many-particle sector
is not devoid of the unresolved problems).
That is, one is presented with an extension of  special relativity
without providing a full replacement of spacetime transformations;
and without providing an operational meaning of the various 
symbols used.
Thus, e.g., the phase space remains either ill defined,
or undefined. So also is the case with the parameter which
attends to the inertial properties of the particles.

Under these circumstances the physical distingushability issue becomes
ill/un-defined and one can only clarify issues which do not invoke 
phase space. One such issue is the question of modification
of Dirac equation, and study of wave equations associated with
different representation spaces.
The task of this section is to show that under these circumstances
one can claim all sort of effects which depend on the deformation 
parameter $\ell_P$, the Planck length. But many of these corrections
carry no operational meaning.
The temptation to claim $\mathcal{O}(\ell_P)$
corrections \cite{Agostini:2002yd}, and even to 
suggest a CPT violation at that order
\cite{Acosta:2004pk}, 
should be resisted and additional conceptual and mathematical questions
asked. 

These claims are now established. In writing these claims
I, by necessity, and to make this work as self-contained as possible,
reproduce some of the results of    
\cite{Ahluwalia-Khalilova:2002wf,Ahluwalia-Khalilova:2002ye,Ahluwalia-Khalilova:2004dc}.
Some of the mathematical aspects are similar to those 
presented by 
Agostini, Amelino-Camelia, and Arzano in
\cite{Agostini:2002yd}
but my interpretation differs dramatically. 
In what follows,
I also incorporate
an important phase factor, not appreciated in  
\cite{Agostini:2002yd}.
The neglect of that phase amounts to projecting out antiparticles from 
\texttt{NSR-2}s as I have already noted in 
\cite{Ahluwalia-Khalilova:2004dc}.

\subsection{NSR-2 of Amelino-Camelia, and
Magueijo and Smolin.}

Simplest of  \texttt{NSR-2}s
result from
keeping the algebra of boost- and 
rotation- generators
intact while modifying the boost parameter in a non-linear 
manner (the deformation of algebra itself lies
in the remaining sector). 
Specifically, in the \texttt{NSR-2} of Amelino-Camelia
the boost parameter, $\bv$, changes from the special relativistic form
\beq
\cosh{\varphi} = \frac{E}{m }\,,\quad
\sinh{\varphi} = \frac{p}{m}\,,\quad
\hbv=\frac{\p}{p}, \label{dirac}
\eeq
to \cite{Amelino-Camelia:2002vy,Bruno:2001mw,Judes:2002bw}

\beq
\cosh\xi &=& 
\frac{1}{\mu}\left( \frac{
e^{\ell_P  E} 
-\cosh\left(\ell_P\, m \right)}
{\ell_P \cosh\left( \ell_P \,m/2\right)}
\right)\,,\label{gac1}\\
\sinh\xi &=& 
\frac{1}{\mu}\left(
\frac{p \,
e^{\ell_P  E}
}
{\cosh\left(\ell_P \,m/2\right)} \right) ,
\quad
\hbxi=\frac{\p}{p}. \label{gac2}
\eeq
While for the \texttt{NSR-2}  of  Magueijo and Smolin the change takes
the form \cite{Magueijo:2001cr,Judes:2002bw}
\beq
\cosh\xi &=& 
\frac{1}{\mu}
\left(\frac{E}{1-\ell_P\,E}\right)
\,,\label{ms1}\\
\sinh\xi &=& 
\frac{1}{\mu}
\left(\frac{p}{1-\ell_P\,E}\right)\,,
\quad
\hbxi=\frac{\p}{p}
\,. \label{ms2}
\eeq
Here,  $\mu$ is a Casimir invariant of \texttt{NSR-2} (see equation 
(\ref{ci}) below) and is given by
\beq
\mu =\cases{
\frac{2}{\ell_P}\,
\sinh\left(\frac{\ell_P \,m}{2}\right) 
                     & $\mbox{for   \cite{Amelino-Camelia:2002vy}'s 
\texttt{NSR-2} }$\\
\frac{m}{1-\ell_P m}
                     & $\mbox{for   \cite{Magueijo:2001cr}'s
\texttt{NSR-2} }$\cr}
\eeq
The notation is that of \cite{Judes:2002bw}; 
with the minor  exceptions:
$\lambda$, $\mu_0$, $m_0$ there are $\ell_P$,  $\mu$, $m$ here. 
Now,  it is an assumption of \texttt{NSR-2} theories that 
the non-linear action of $\bx$ is restricted to the momentum space 
\textit{only}. No fully satisfactory spacetime description
in the context of the \texttt{NSR-2} theories has yet emerged, and 
we are not sure if such an operationally meaningful 
description indeed exists. Therefore, to the extent
possible, our arguments shall be confined to the momentum space.

\subsection{Master equation for spin-1/2: Dirac case.}

Since the relevant underlying 
spacetime symmetry generators remain unchanged much of the  
formal apparatus of the finite-dimensional representation
spaces associated with the Lorentz group remains intact.
In particular, there still exist $(1/2,\,0)$ and $(0,\,1/2)$
spinors. But now they transform from the rest frame to
an inertial frame in which the particle has momentum, $\p$, as
\beq
\phi_{(1/2,\,0)}\left(\p\right)
& = & \exp\left( + \frac{\bs}{2}\cdot\bx \right)
\phi_{(1/2,0)}\left(\0\right)\,,\label{a}\\
\phi_{(0,\,1/2)}\left(\p\right)
& = & \exp\left(- \frac{\bs}{2}\cdot\bx \right)
\phi_{(0,1/2)}\left(\0\right)\,.\label{b}
\eeq  
Since 
the null momentum vector $\0$ is still isotropic,
one may assume that (see p 44  of 
\cite{Ryder} \textit{and}  
\cite{Ahluwalia:1993zt,dva_review,Gaioli:1998ra}): 
\beq
\phi_{(0,1/2)}\left(\0\right) = \zeta \,\phi_{(1/2,0)}\left(\0\right)\,,
\label{c}
\eeq
where $\zeta$ is an undetermined phase factor. 
In general, the phase
$\zeta$ encodes
C, P, and T properties. The interplay of
equations (\ref{a})-(\ref{c}) yields a master equation for
the $(1/2,\,0)\oplus(0,\,1/2)$ spinors,
\beq
\psi\left(\p\right) = \left(
			\begin{array}{c}
			\phi_{(1/2,\,0)}\left(\p\right)\\
			\phi_{(0,\,1/2)}\left(\p\right)
			\end{array}
		     \right)\,,
\eeq
to be
\beq
\left(
\begin{array}{cc}
-\zeta \1_2 & \exp\left(\bs\cdot\bx\right) \\
\exp\left(- \bs\cdot\bx\right) & - \zeta^{-1} \1_2
\end{array}
\right) \psi\left(\p\right) = 0\,,\label{meq}
\eeq
where $\1_n$ stands for $n\times n$ identity matrix
(and $0_n$  represents the corresponding null matrix).
As a check, taking $\bx$ to be $\bv$, and after some
simple algebraic manipulations,
the master equation (\ref{meq}) reduces to
\beq
\left(
 	\begin{array}{cc}
	- m \zeta \1_2 & E \1_2 + \bs\cdot \p \\
	E \1_2 - \bs\cdot \p & - m \zeta^{-1} \1_2
	\end{array} 
\right) \,
\psi\left(\p\right) = 0\,.\label{d}
\eeq
With the given identification of the boost parameter
we are in the realm
of special relativity. There, the operation of
parity is well understood. Demanding  parity
covariance for equation (\ref{d}), we obtain
$\zeta=\pm 1$. Identifying 
\beq
\left(  \begin{array}{cc}
	0_2 & \1_2 \\
	\1_2 & 0_2
	\end{array}
\right)\,,\quad
\left(  \begin{array}{cc}
	0_2 & -\bs\\
	\bs & 0_2
	\end{array}
\right)\,,
\eeq
with the Weyl-representation $\gamma^0$, and $\gamma^i$, respectively;
equation (\ref{d}) reduces to the Dirac equation of  special relativity,
\beq
\left(\gamma^\mu p_\mu \mp m\right)\psi\left(\p\right)=0\,.\label{de}
\eeq
The linearity of the Dirac equation in $p_\mu= (E,-\p)$, is now clearly
seen to be associated with two observations: 

\begin{enumerate}
\item[$\mathcal{O}_1$.]
that, $\bs^2 = \1_2$; and
\item[$\mathcal{O}_2$.]
that in special relativity, the hyperbolic functions
\textendash~ 
see equation (\ref{dirac}) \textendash~ associated with the boost parameter
are linear in $p_\mu$. 
\end{enumerate}
In \texttt{NSR-2}, observation $\mathcal{O}_1$
still holds. But, as equations (\ref{gac1})-(\ref{ms2}) 
show, $\mathcal{O}_2$ is strongly violated.
The extension of the presented formalism to 
the eigenspinors of the charge conjugation 
operator is more subtle 
\cite{Ahluwalia-Khalilova:2004ab}.
The extension to $(1/2,1/2)$ representation space to describe 
vector particles is less demanding and can be immediately 
obtained using the techniques of 
\cite{Ahluwalia:2000pj}\footnote[1]{Such an exercise has 
been undertaken in \cite{Acosta:2004pk}, but it suffers from 
a set of serious interpretational 
issues. A matter on which I shall briefly comment
in this  paper.
For the moment the reader is warned that what authors
of \cite{Acosta:2004pk} call helicity is really
spin projection on the $\widehat{z}$ direction.
This already introduces several errors as the $\p$
vector in \cite{Acosta:2004pk} is a completely general
special-relativistic three momentum.}.

\subsection{Master equation for higher spins.}

The above-outlined procedure applies
to all, bosonic as well as fermionic,  
$(j,0)\oplus(0,j)$ representation spaces. It is 
not confined to $j=1/2$. A straightforward generalization 
of the $j=1/2$ analysis immediately yields the Master equation 
for an arbitrary-spin,
\beq
\left(
\begin{array}{cc}
-\zeta\, \1_{2j+1}& \exp\left(2\J\cdot\bx\right) \\
\exp\left(- 2\J\cdot\bx\right) & - \zeta^{-1} \, \1_{2j+1}
\end{array}
\right) \psi\left(\p\right) = 0\,,\label{j}
\eeq
where
\beq
\psi\left(\p\right) = \left(
			\begin{array}{c}
			\phi_{(j,\,0)}\left(\p\right)\\
			\phi_{(0,\,j)}\left(\p\right)
			\end{array}
		     \right)\,.\label{js}
\eeq
Equation (\ref{j}) 
contains the central result of the previous section as a 
special case.  
For studying the special relativistic limit it is convenient 
to bifurcate the
$(j,0)\oplus(0,j)$ space into two sectors by splitting the 
$2(2j+1)$ phases, $\zeta$,
into two sets: $(2j+1)$ phases $\zeta_+$, and the other  
$(2j+1)$ phases $\zeta_-$. Then  in particle's rest frame
the $\psi(\p)$ may be written as
\beq
\psi_h(\0)=
\cases{
u_h(\0) &  $\mbox{when}~ \zeta=\zeta_+$\cr
v_h(\0) &  $\mbox{when} ~\zeta=\zeta_- $\cr
}
\eeq
The explicit forms of $u_h(\0)$ and $v_h(\0)$ (see equation (\ref{c})) are:
\beq
u_h(0)=\left(
\begin{array}{c}
\phi_h(\0) \\
\zeta_+ \,\phi_h(\0)   
\end{array}
\right),\,
v_h(0)=
\left(
\begin{array}{c}
\phi_h(\0) \\
\zeta_-\, \phi_h(\0)   
\end{array}
\right),
\eeq
where the   $\phi_h(\0)$ are defined as
$\J\cdot \hp\, \phi_h(\0) = h \,\phi_h(\0)$, and $h=-j,-j+1,\ldots,+j$.
In the parity covariant special relativistic 
limit, we find $\zeta_+ = +1$ while   
$\zeta_- = -1$.
As a check, for $j=1$, identification of $\bx$ with $\bv$,
and after implementing parity covariance, equation (\ref{j}) yields
\beq
\left(\gamma^{\mu\nu}p_\mu p_\nu \mp m^2\right)\psi(\p)=0\,.\label{bwweq}
\eeq 
The  $\gamma^{\mu\nu}$ are unitarily equivalent  
to those of \cite{Ahluwalia:1993zt}, and thus we reproduce 
\textit{bosonic matter fields}
with $\left\{C,\,P\right\} = 0$.  A carefully taken massless limit then shows
that the resulting equation is consistent with the free Maxwell equations
of electrodynamics.
This again casts doubts
on the operational distinguishability of
the $\mathcal{O}(\ell_P)$ predictions presented in
\cite{Acosta:2004pk}.

Since the $j=1/2$ and $j=1$ representation spaces of \texttt{NSR-2} reduce to
the Dirac and Maxwell descriptions, it would seem apparent
(and as is
often argued in similar contexts \cite{Agostini:2002yd} 
\textendash~ wrongly, at least to an extent, 
as we will soon see)  that
the \texttt{NSR-2}
contains physics beyond the linear-group realizations of special relativity.
To the lowest order in $\ell_P$,  equation (\ref{meq}) yields
\beq
\left(
\gamma^\mu {p}_\mu + \tilde{m} + 
\delta_1\, \ell_P\right)
\psi(\p)=0\,,
\eeq
where
\beq
\tilde{m}
&=& \left(
\begin{array}{cc}
-\zeta \1_2  & 0_2  \\
0_2 & -\zeta^{-1} \1_2
\end{array}
\right)\,m \,
\eeq
and
\beq
\delta_1 =
\cases{
\gamma^0\left(\frac{E^2-m^2}{2}\right)+\gamma^i p_i\, E
& $\mbox{for  \cite{Amelino-Camelia:2002vy}'s \texttt{NSR-2}}$\\
\gamma^\mu p_\mu\, \left(E-m\right)
&       $\mbox{for  \cite{Magueijo:2001cr}'s \texttt{NSR-2}}$
}
\eeq
Similarly, the presented master equation can be used  
to obtain \texttt{NSR-2}'s counterparts for Maxwell's electrodynamic.
Unlike the Coleman-Glashow framework \cite{Coleman:1998ti}, 
the existing \texttt{NSR-2}s 
provide \textit{all} 
corrections, say, to the standard model of the high energy physics, 
in terms of \textit{one} 
\textendash~ and \textit{not} $46$ \textendash~ 
fundamental constant, $\ell_P$.
Had \texttt{NSR-2}s been operationally well-defined and 
distinct this would have been a
remarkable power of \texttt{NSR-2}-motivated frameworks.

\subsection{
Judes-Visser 
variables: challenging some of the \texttt{NSR-2}'s claims}

We now show that the  \texttt{NSR-2} program as implemented currently is 
inadvertently misleading. Sometimes this is in good humor (or, so
we interpret), e.g. when the author of \cite{Magueijo:2003gj} notes, 
`mathematical triviality by no means implies physical equivalence, 
and one may argue that it is in fact an asset'.
At other times, 
it is simply a manifest lack of care being given
to the operational meaning of various symbols one uses in his/her
mathematical formalism and a total disregard for the existing literature
on `indistinguishability', or `conceptual' issues for \texttt{NSR-2} 
\cite{Acosta:2004pk}.

The question is what are the operationally measurable
quantities in  \texttt{NSR-2}? The $E$ is no longer the $0$th component, 
nor is $\p$ the spatial component  of $4-$momentum. 
Neither  is $m$   an invariant under the  \texttt{NSR-2} boosts. 
Their physical 
counterparts, as we interpret them, are  
Judes-Visser  variables \cite{Judes:2002bw}, $\eta^\mu \equiv 
\left(\epsilon(E,p),\,\bp(E,p)\right)=(\eta^0,\be)$, and $\mu$.
The $\epsilon(E,p)$ and $\bp(E,p)$ 
relate to the rapidity parameter $\bx$ of  \texttt{NSR-2}
in the same functional form
as do $E$ and $\p$ to  $\bv$ of special relativity:
\beq
\cosh\left(\xi\right)= \frac{\epsilon(E,p)}{\mu}\,,\quad
\sinh\left(\xi\right)= \frac{\pi(E,p)}{\mu}\,,
\eeq
where 
\beq
\mu^2= \left[\epsilon(E,p)\right]^2 - \left[\bp(E,p)\right]^2\,.
\label{ci}
\eeq
They provide 
the most economical and physically transparent formalism for 
representation space theory in \texttt{NSR-2}.
For $j=1/2$ and $j=1$, equation (\ref{j}) yields the \textit{exact
NSR-2} equations for $\psi(\bp)$:
\beq
\hspace{-36pt}\left(\gamma^\mu \eta_\mu + \tilde{\mu}\right)
\psi\left(\be\right)=0\,, \quad{\mbox{where}}\;\;
\tilde{\mu}
= \left(
\begin{array}{cc}
-\zeta^{-1} \1_2 & 0_2 \\
0_2 & -\zeta \1_2
\end{array}
\right)\,\mu\,, &&                \label{denew}\\
\hspace{-36pt}\left(\gamma^{\mu\nu}\eta_\mu \eta_\nu + \tilde{\mu}^2\right)
\psi(\be)=0\,,\quad{\mbox{with}}\;\;
\tilde{\mu}^2
= \left(
\begin{array}{cc}
-\zeta^{-1} \1_3 & 0_3 \\
0_3 & -\zeta \1_3
\end{array}
\right)\,\mu^2\,.&&
\label{bwweqnew}
\eeq
From an operational point of view the $\eta^\mu$ and $\mu$
\textit{are} the physical observables. The old operational meaning of
the symbols $E$ and $\p$ is lost in the non-linear realization
of the boost in the momentum space\footnote{The adherents of
\texttt{NSR-2} propose $p^\mu$ as a \textit{physical}
variable despite the fact that it has no known geometrical
object that it can be identified with; see, 
\cite{Liberati:2004ju} for a detailed discussion of
interpretational possibilities of this view and the 
questions it raises. On the other hand, 
the Judes-Visser variables $\eta^\mu$ have well-defined mathematical
meaning for all inertial observers (without requiring a preferred 
observer) and correspond to conserved quantities. The  
\texttt{NSR-2} view that `measured quantities $p^\mu$' 
need not be conserved, considered as a possibility in
  \cite{Liberati:2004ju} to lend 
\texttt{NSR-2} an element of viability, amounts to
giving a  non-linear choice of coordinates in
momentum space a far fetched, and unviable, 
physical interpretation. That is, I do not accept the
plausibility argument of  \cite{Liberati:2004ju} that
`unscreenable' quantum gravitational effects may
forbid $\eta^\mu$ from being directly measured.
One reason for this stance is that the `unavoidable'
quantum-mechanical fluctuations do not make 
the energy-momentum $4$-vector as immeasurable; in fact,
quantum framework provides precise calculational tools
to predict the associated uncertainties. Had this not been
the case Dirac equation would have not carried the 
empirically-verified (and demanded by the Lorentz covariance)
linearity in time acquired via the identification 
$p_\mu \,\to\,i\partial_\mu$.}.
There is
covariance of the form of the considered\footnote[3]{The result
is expected to be the same for other representation spaces.} fermionic 
and bosonic wave equations 
under the transformations
\beq 
m \rightarrow \mu\,,\quad p^\mu \rightarrow \eta^\mu\,,\quad
\bv \rightarrow \bx\,. &&
\eeq
Thus, at the level of momentum-space wave equations, 
the \texttt{NSR-2} and  
special relativistic descriptions of fermions and bosons
carry identical forms. For this reason, the 
 $\mathcal{O}(\ell_P)$ departures 
given in  \cite{Agostini:2002yd} are artifacts 
of the used variables.
The same holds true for the  $\mathcal{O}(\ell_P)$ corrections 
at the level
of vector potential given in  \cite{Acosta:2004pk}.
Again, they are simply artifacts of the used variables.
To be more precise, the object considered by authors
of \cite{Acosta:2004pk}
Lorentz transforms as $(1/2,1/2)$, and is represented as 
$\mathcal{A}^\mu(\p)$ in  \cite{Acosta:2004pk}.
On the other hand the related field strength tensor,
$\mathcal{F}^{\mu\nu}(\p)$, transforms as $(1,0)\oplus(0,1)$
object under Lorentz transformations. Since the latter
representation space in the momentum space is 
shown here to be indistinguishable
from its special-relativistic counterpart
\textemdash~ see, equation (\ref{bwweqnew2}) \textemdash~ 
the associated \texttt{NSR-2}'s  $\mathcal{A}^\mu(\p)$
cannot be physically different from its special-relativistic 
counterpart. Thus, it establishes that the Acosta-Kirchbach
result on  $\mathcal{O}(\ell_P)$ departures is  an artefact
of ignoring Judes-Visser variables. 

Yet, it should be noted that to the extent the phase
space of \texttt{NSR-2} remains ill/un-defined one 
cannot make any physically-significant ``indistinguishability'' 
claim at the level of physical amplitudes and cross sections in 
\texttt{NSR-2}. Same remains true for \texttt{NSR-3}.

Given that so far a satisfactory spacetime theory 
of \texttt{NSR-2} is lacking, we implement parity covariance 
by demanding covariance under $\be \rightarrow \be^\prime \defn -\be$.
This demand, transforms the above wave equations to
\beq
&& \left(\gamma^\mu \eta_\mu \mp {\mu}\right)
\psi\left(\be\right)=0  \,,           \label{denew2}\\
&& \left(\gamma^{\mu\nu}\eta_\mu \eta_\nu \mp {\mu}^2\right)
\psi(\be)=0\,.
\label{bwweqnew2}
\eeq
The translation of equations (\ref{denew2}) to (\ref{bwweqnew2}) to
spacetime occurs as follows. Once $\eta^\mu$ are accepted
as physical, they immediately require the spacetime
variables to be that of standard special relativity; otherwise,
there is no meaning to the interpretation of  $\eta^\mu$ 
as corresponding  to the measured and conserved energy-momentum
$4$-vector  implied by time-translational invariance. 
The spacetime evolution follows with the substitution
$\eta_\mu \,\to\,i\partial_\mu$, and 
$\psi(\be) \,\to \exp(\mp i \eta_\mu x^\mu)\psi(\x)$.

The basic questions which the discussion of  Acosta and Kirchbach paper
\cite{Acosta:2004pk} now provokes are
\begin{enumerate}
\item[$\bullet~$]
as to which parameter $m$, or $\mu$,
is connected with the CPT symmetry of the 
 2-scale non-linear deformations
of special relativities of 
\cite{Amelino-Camelia:2000mn,Magueijo:2001cr}
\item[$\bullet~$]
 and, which parameter,  $m$, or $\mu$,
in the non-relativistic Ehrenfest, weak-field, limit
of the  2-scale non-linear deformations
of special relativities of 
\cite{Amelino-Camelia:2000mn,Magueijo:2001cr}
couples to gravity
\end{enumerate}
The authors of \cite{Acosta:2004pk} do not ask these, or related, 
questions.
Barring
certain pathologies of the theories described in 
\cite{Amelino-Camelia:2000mn,Magueijo:2001cr},
the answer to the question must be  independent of the representation
space.
 Therefore, it is sufficient
to establish it for equation (\ref{denew2}), i.e., 
the  $\left(\frac{1}{2},0\right)\oplus \left(0,\frac{1}{2}\right)$ 
representation space. The answer constitutes nothing more than a 
simple and obvious exercise. Yet, to lay matters to rest, we
outline the details. 

First, we know from the above discussion, and further details
given in  
\cite{Ahluwalia-Khalilova:2004dc},
that in the Weyl representation the particle 
spinors read
\beq
u_\pm\left(\be\right) = 
\kappa^{\left(\frac{1}{2},0\right)\oplus \left(0,\frac{1}{2}\right)}
\left(
			\begin{array}{c}
			\phi_\pm\left(\0\right)\\
			+\,\phi_\pm\left(\0\right)
			\end{array}
		     \right)\,,
\eeq
while the antiparticle spinors are

\beq
v_\pm\left(\be\right) = 
\kappa^{\left(\frac{1}{2},0\right)\oplus \left(0,\frac{1}{2}\right)}
\left(
			\begin{array}{c}
			\phi_\pm\left(\0\right)\\
			-\,\phi_\pm\left(\0\right)
			\end{array}
		     \right)\,.
\eeq
The ${\left(\frac{1}{2},0\right)\oplus \left(0,\frac{1}{2}\right)}$
boost operator which appears in the above equations is
\beq
\kappa^{\left(\frac{1}{2},0\right)\oplus \left(0,\frac{1}{2}\right)}
\defn
\left(
\begin{array}{cc}
\exp\left( + \frac{\bs}{2}\cdot\bx \right) & \0_2 \\
\0_2 & \exp\left( - \frac{\bs}{2}\cdot\bx \right)
\end{array}
\right)\,,
\eeq
and $\phi_\pm\left(\0\right)$ are eigenspinors of 
the helicity operator $\frac{\bs}{2}\cdot\hbxi $.
These are connected by the charge conjugation symmetry
only if inertial properties of the 
described particles are dictated by $\mu$, and not $m$.
Second, taking the non-relativistic Ehrenfest limit of equation (\ref{denew2})
in the presence weak gravity reveals that it is $\mu$ that couples to
gravity and not $m$.

The answer for both of the above-asked questions is, $\mu$. Not, $m$.
The $m$ is simply not a physically-meaningful  observable of the theories
proposed in \cite{Amelino-Camelia:2000mn,Magueijo:2001cr}, and 
\cite{Kowalski-Glikman:2004kp}.
The asymmetry noted by Acosta and Kirchbach is superficial. It occurs
only if one ignores the asked questions. As such, at this stage of analysis,
there is no argument which favours  CPT violation.
Equations, such as (11), of  \cite{Acosta:2004pk} are thus devoid of 
any physical content. They arise on one hand 
due to misidentification of the inertial properties of particles
and on the other due to implicit neglect of 
Judes-Visser  transformations \cite{Judes:2002bw}.

The next level of clarification that is needed is that the apparent
similarity between the equaqtions (\ref{denew2}) and (\ref{bwweqnew2})  
and their special relativistic counterparts
can make one conclude that there is no distinction
between \texttt{NSR-2}s, or  \texttt{NSR-3} (for in the latter
too the Lorentz algebra remains undeformed). Such a claim, besides
already noted reasons, would  
appear to be misleading, because while 
in ordinary special relativity 
the commutator $[P_\mu,P_\nu]$  vanishes, so is not the case
for  
\texttt{NSR-2}s and  \texttt{NSR-3}. However, 
Chryssomalakos and Okon show \cite{Chryssomalakos:2004wc}
that the algebra of 
\texttt{NSR-3} can be brought to Lie, i.e. linear, form by a correct
identification of its generators. If one complements the 
Chryssomalakos-Okon result with the observation that
if one tentatively \textit{defines}\footnote[4]{A matter with
which I'll take issue below.}
 special-relativistic  (1, 2, or 3
invariant scales are irrelevant, but only `special' is relevant) 
theory as the one which carries kinematical group of the tangent
space to the spacetime manifold
one must, in notation of Chryssomalakos and Okon 
 \cite{Chryssomalakos:2004wc}, take $R\rightarrow \infty$ 
limit of the algebra. In that limit the  
$[P_\mu,P_\nu]$  vanishes again. So, in momentum space
equations (\ref{denew2}) and (\ref{bwweqnew2})  
and their special relativistic counterparts are identical.

Yet, the indistinguishability issue cannot be settled
on this last remark alone. The reason is that 
the $R\rightarrow \infty$ limit
still leaves the underlying spacetime noncommutative. 
However, as Mendes has shown \cite{VilelaMendes:1994zg} the spacetime
representation of the $p_\mu$ is still 
$i\partial_\mu$. This implies that the configuration space 
form of equations (\ref{denew2}) and (\ref{bwweqnew2})  
remains the same as in the ordinary special relativity.
Yet, the extension of special relativities based
on Lie-algebraic 
deformations are profoundly different form ordinary
special-relativistic theories. 
The reason is that the Yang-Mendes algebra brought to
attention in Chryssomalakos-Okon's paper \cite{Chryssomalakos:2004wc}
carries a different phases space
\cite{VilelaMendes:2004up}
 and that has the potential to dramatically change
the predictions of the theory at the Planck scale.

We conclude, echoing again the sentiments of 
\cite{Chryssomalakos:2004wc}, 
that the search for relativities, special or general,
with additional invariant scales while well motivated,
carries no justification to go towards non-linear 
extensions. These \texttt{NSR}s leave far too many questions unanswered
with dubious justification for considering them in the first 
place. At this early stage there is no reason to invoke 
non-linear deformations. 
On the other hand,
the Lie, i.e. linear, deformations are well-defined.
The stabilized Poincar\'e-Heisenberg algebra is 
nothing but Yang-Mendes Lie algebra 
\cite{VilelaMendes:1994zg,Yang:1947ud}. 
Special and general relativities with additional
invariant scales, in my opinion,
must be based on this yet-unexplored Lie algebraic structure.
In any case, despite differences of opinions, one thing is
clear that deformation of spacetime symmetries involving 
$c$, $\hbar$, $\ell_P = \sqrt{\hbar G/c^3}$ (which adds to $c$ and $\hbar$
 the constant $G$), and 
the cosmological  
constant $\Lambda$ \textemdash~ or, appropriate new combinations 
\textemdash~ 
shall play a profound role in any theory of quantum gravity
\cite{Amelino-Camelia:2003xp,Ahluwalia-Khalilova:2004dc} and that it
may already be evident
in certain anomalous astrophysical and cosmological observations  
\cite{Kowalski-Glikman:2004kp}.
One caution should, however, be exercised: there is a tendency
in the literature, see, e.g.,    \cite{Freidel:2003sp},
to identify modification of the certain algebraic commutators,
or deformation of dispersions relations, with 
\texttt{NSR}-n's.  Such naive identifications confuse the
issue of non-linear versus linear (i.e. Lie) deformations.
Not all modifications which lead to such departures
from special relativity fall under the umbrella of 
\texttt{NSR}-n's. In fact, as we shall momentarily see,
\texttt{NSR}-n's probably are not viable physical theories.
Yet, 
more viable theories based on Lie-algebraic structures
carry spacetime/energy-momentum non-commutativity,
and many other intrinsic features similar, though not the same, as to 
those found in \texttt{NSR}-n's.

\section{A stable Lie-algebraic structure 
for freely falling frames at the interface of 
gravitational and quantum realms}
\label{Sec:FreeFall}

Mendes \cite{VilelaMendes:1994zg} and 
{Chryssomalakos \cite{Chryssomalakos:2001nd} have emphasized that
stability of the underlying Lie algebras must be taken as one of the
important physical criterion to consider a theory 
as  physically viable.
In the context of this newly suggested criterion, 
if one confines only to spacetime symmetries
then one notes that the algebra underlying special relativity,
i.e., Poincar\'e, is stable modulo the remarks 
made earlier.
However, as soon as quantum phenomena are studied one must 
bring in Heisenberg algebra, which is also stable (up to
a `harmless' instability noted earlier), into the picture. 
But the Poincar\'e-Heisenberg algebra
ceases to be stable. These  observations may be further 
amplified by noting:

\begin{enumerate}
\item[$\bullet~$]
In 1947 Snyder pointed out that the assumption that spacetime 
be a continuum is not imposed by  Lorentz invariance \cite{Snyder:1946qz}.

\item[$\bullet~$]
Later in the same year, Yang noted that lack of translational invariance
in Snyder's framework can be rectified if spacetime is allowed to carry
curvature \cite{Yang:1947ud}. In that same one-and-half column 
paper, Yang also presented the complete Lie algebra associated with 
the suggested modification.

\item[$\bullet~$]
In a series of paper published in the last decade (and which
remain almost unnoticed, see, 
e.g.,  \cite{VilelaMendes:1994zg,VilelaMendes:1996pla}),
Mendes 
came to the conclusion that when Poincar\'e and Heisenberg algebras
are considered together \textemdash~ as they must be in any
relativistic quantum framework  \textemdash~ they are not rigid
in the mathematical sense. That is, the physical theories
based on them lack certain elements of robustness, or stability.
In addition, he obtained the suggested stable Lie algebra.
That Lie algebra contained two additional length scales, and it
was the same very algebra as was obtained by Yang in 
1947. Not only this, in support of his suggestion
that stability should be considered an important physical criterion 
he pointed out that both the `relativistic revolution' and the
`quantum revolution' of the last century can be motivated \textemdash~
alas in retrospect
\textemdash~ by the stability criterion.\footnote[6]{When Mendes 
wrote his papers it appears that while
he was aware of Snyder's paper, Yang's important paper had 
escaped his attention.}

\item[$\bullet~$]
Taking note of the work by Mendes,  Chryssomalakos and Okon have just noted
that recently proposed triply-special-relativity algebra 
proposed by Kowalski-Glikman and Smolin \cite{Kowalski-Glikman:2004kp}
can be brought to a linear (i.e., Lie) form by a correct 
identification of its generators and that
the resulting Lie algebra is precisely of Yang-Mendes form.

\end{enumerate}

As, in essence, the whole story began with Snyder, one is tempted to
suggest the stabilized form of
Poincar\'e-Heisenberg algebra, i.e. Snyder-Yang-Mendes algebra, 
be called by a simple acronym \texttt{SYM}. But this acronym has already
been used widely in the literature in other contexts, such as for
super Yang-Mills theories, and for that reason we shall 
settle for `Lie algebra for \texttt{IGQR}'.

Referring to the
\textit{Introduction}, one can now identify the two length scales that
appear in the Lie algebra for \texttt{IGQR} 
with $\ell_P$ and $\ell_C$. With the indicated identifications,
the Mendes-inspired work of Chryssomalakos and Okon 
\cite{VilelaMendes:1994zg,Chryssomalakos:2004gk} 
suggests

\beq
&& \left[J_{\mu\nu},J_{\rho\sigma}\right] = 
i \left(
 \eta_{\nu\rho} J_{\mu\sigma}+\eta_{\mu\sigma} J_{\nu\rho}  
 - \eta_{\mu\rho} J_{\nu\sigma}  
- \eta_{\nu\sigma} J_{\mu\rho} \right)\,,\label{eq:a1}\\
&& \left[J_{\mu\nu}, P_\lambda\right] = i \left(\eta_{\nu\lambda} P_\mu 
- \eta_{\mu\lambda} P_\nu \right)\,,\\
&& \left[J_{\mu\nu}, X_\lambda\right] = i \left(\eta_{\nu\lambda} X_\mu 
- \eta_{\mu\lambda} X_\nu \right)\,,\\ \nonumber\\
&& \left[P_\mu,P_\nu\right] = 
 i \frac{\hbar^2}{\ell^2_C} J_{\mu\nu}\,,\\
&& \left[X_\mu,X_\nu\right] =  i {\ell^2_P} J_{\mu\nu}\,, \label{eq:ncst}\\
&& \left[P_\mu,X_\nu\right] = 
i  \hbar \eta_{\mu\nu} \mathcal{F}  + i \hbar  \beta\, 
J_{\mu\nu}\,,\label{eq:hfc}\\ \nonumber\\
&& \left[P_\mu,\mathcal{F}\right] =
 i \frac{\hbar}{\ell^2_C} X_{\mu} - i \beta P_\mu \,,\\
&& \left[X_\mu,\mathcal{F}\right] = i \beta X_\mu 
- i \frac{\ell^2_P}{\hbar} P_{\mu}
\,,\\
&& \left[J_{\mu\nu},\mathcal{F}\right] = 0\,,\label{eq:a2}
\eeq 
as a natural candidate for a physically viable theory 
in the \texttt{IGQR}. Here $\beta \in {R}$ is a \textit{new} 
dimensionless constant. Its presence has
been noted in  
\cite{VilelaMendes:1994zg,Chryssomalakos:2004gk,Khruschev:2002cq}
with differing emphasis. In
this section we exhibit $c$ and $\hbar$ explicitly.
As in the previous section, $\eta_{\mu\nu}$ is diagonal with
diag$(1,-1,-1,-1)$; whereas $p^\mu = (E/c,\p)$ and $p_\mu = (E/c, -\p)$.

There is now a temptation to take $\ell_C \to \infty$ limit and 
identify the resulting algebra with the algebra that shall be found
in freely falling frames of quantum gravity. 
A hint in that direction occurs in papers of Mendes.
This, in my opinion, may \textit{not} 
be justified as then owing to work of Yang \cite{Yang:1947ud}
translational invariance is no longer obvious (a question
which should be re-examined in any case for $\beta\ne 0$). 
But, more importantly,
the modified zero point energy \textemdash~ which now need
not carry the same magnitude for fermionic and bosonic fields 
\textemdash~ cannot be removed from a freely falling frame. This
is as true for \texttt{IGQR} as for freely falling frames of
special relativity endowed with any quantum field. The difference now
is that the magnitude of the zero point energy at a given angular
frequency is not guaranteed to be equal (a subject which
requires \textit{ab initio} calculations). It may indeed happen that
the famous `120 orders of magnitude problem' associated with the
cosmological context in the standard Poincar\'e-Heisenberg framework
is resolved  by cancellation of the zeroth order 
`$\pm \,(1/2)\, \hbar \omega$' contributions, and the observed cosmological 
constant arises due to  higher order terms. However, for
such a cancellation to occur supersymmetry would seem most natural
agent. But such a circumstance
would require 
a non-trivial stability analysis with supersymmetry incorporated. 
This is not to be interpreted as a weakness of the Lie-algebraic 
stability paradigm but as it strength because it suggests 
a logical and well-defined path to be followed.
Also if above-mentioned translational invariance is lost,
operational meaning for energy and momentum ceases to exist.

For these reasons, 
I suggest that the Lie algebra for \texttt{IGQR} as it is written above
represents the algebra of new special relativity that underlies
the \texttt{IGQR}.
 It has an intrinsic curvature. That is,  the \texttt{IGQR} spacetime
carries a curvature even in the absence of conventional sources. The source of
this curvature is the  vacuum energy density that defines 
the cosmological constant
and cannot be eliminated from freely falling frames as can be justified 
on empirical grounds also. 
In this interpretation, quantum gravity is likely to arise 
from `gauging' this algebra in precisely the same sense as in Yang-Mills 
gauge theories. In such a theory the notion of point particle
is no longer a viable one. It is replaced by a fuzzy specification
governed by spacetime noncommutativity (\ref{eq:ncst}).
 Similarly,
de Broglie wave-particle duality suffers a modification due to
deformation of the fundamental commutator (see below, again). The locality 
of the standard quantum field theory (which is based on
unstable Poincar\'e-Heisenberg algebra) is lost; see equation 
(\ref{eq:ncst}). This latter
unavoidable feature is likely to introduce an intrinsic element
of CPT violation.

The fact that the Heisenberg's fundamental
commutator (\ref{eq:hfc}) undergoes  non-trivial modifications
with $\mathcal{F}$ ceasing to be central, and $\beta \ne 0$, has the
following
immediately identifiable 
consequence:
the position-momentum Heisenberg uncertainty 
relations get modified. For example,
\beq
\Delta x \,\Delta p_x  \ge \frac{\hbar}{2}
\left\vert \left\langle 
  \mathcal{F} 
\right\rangle\right\vert\,,
\eeq
 while $\Delta x\, \Delta p_y $ no longer vanishes, but instead is
given by
\beq
\Delta x\, \Delta p_y  \ge \frac{\beta\hbar}{2}
\left\vert \left\langle 
  J_z 
\right\rangle\right\vert\,.
\eeq
That is, $\Delta x \,\Delta p_x$ is sensitive
to $\mathcal{F}$; while sensitivity to $\beta$ is
carried in $\Delta x\, \Delta p_y$.
Furthermore, in the usual notation, one has the following 
representative expression for the product of uncertainties
in position measurements:
\beq
\Delta x\, \Delta y  \ge \frac{\ell_P^2}{2}
\left\vert \left\langle 
  J_z 
\right\rangle\right\vert\,,\label{eq:xy}
\eeq
with
\beq
\Delta p_x\, \Delta p_y  \ge \frac{\hbar^2}{2\ell_C^2}
\left\vert \left\langle 
  J_z 
\right\rangle\right\vert\,.
\eeq
complementing equation (\ref{eq:xy}) for momentum measurements 
(the expectation value, denoted by $\langle\ldots\rangle$
in the above expressions, is with respect states that arise
in a (yet to be fully formulated) quantum field theory based on
Lie algebra for \texttt{IGQR}).
In addition
\begin{quote}
\begin{enumerate}
\item[$\bullet~$]
The de Broglie wave-particle duality must undergo a 
profound modification.
\item[$\bullet~$]
The notion and magnitude 
of zero-point energy must suffer corrections. 
\item[$\bullet~$]
 The equal-energy spacing of vibrational and rotational 
diatomic states must undergo well-defined corrections.
\item[$\bullet~$]
The concept of \textit{particle} is now defined via the Casimir
invariants for  Lie algebra for \texttt{IGQR}.
\item[$\bullet~$]
The quantization procedure require an \textit{ab initio}
formulation where the demand of locality is abandoned.
\end{enumerate}
 \end{quote}
This circumstance immediately asks for a detailed investigation
of these modifications/corrections not only to examine the
possibility of experimental confirmation of the suggested
Lie algebra by precision experiments
but also to place limits on $\beta$ from existing data.
The last two inferences are directly related. For 
$\beta=0$, they are confirmed by Mendes \cite{VilelaMendes:1999xv}. 
An examination of momentum-space wave equations along the lines
presented in section \ref{Sec:Proverbial}, with explicit modification
to `$p_\mu \,\to\,i\hbar \partial_\mu$' consistent with the 
algebra (\ref{eq:a1})-(\ref{eq:a2}), 
should yield $\beta,\ell_P,\;\mbox{and}\;\ell_C$ 
dependent corrections to standard model physics while at the same time 
defining any possible modification to 
the principle of equivalence\footnote{In this
context a parenthetic observation may be made that now the 
notion of inertial mass undergoes an unavoidable change as 
$P_\mu P^\mu$ is no longer the Casimir for the 
(\ref{eq:a1})-(\ref{eq:a2}).}. 
This exercise should also allow us to study possible violation of CPT
induced by Lie algebra for \texttt{IGQR}. Apart from
carrying intrinsic worth, the result on possible CPT violation
carries enormous
relevance to the LSND excess event anomaly 
\cite{Ahluwalia:1998xb,Murayama:2000hm} and would be most natural
explanation if MiniBOONE confirms LSND result \cite{Ray:2004hp}.

These are all welcome features, which must be investigated in the new
context of Lie algebra for \texttt{IGQR}. In principle, any deviations 
from the standard 
zero-point energy can be studied by precision 
laboratory experiments involving 
\textit{Planck-mass} oscillators 
of superconducting quantum interference devices (SQUIDs).
This can be readily seen from the fact that SQUIDs carry
superconducting currents with \textit{temperature-tunable} 
superconducting mass
\beq
m_s(T) \sim f(T)\, N_a\, m_c\,,
\eeq
behaving as \textit{one} quantum object. In the above equation,
$N_a \approx 6 \times 10^{23}\;\mbox{mole}^{-1}$, $m_c
\approx 2 \times 0.9\times 10^{-27}\;\mbox{gm}$, and $f(T)$
encodes the fraction of the available electrons that are in
a superconducting Cooper state at temperature, $T$. Sufficiently
below the critical temperature, $f(T)$ may approach unity, thus allowing
$m_s(T)$ to easily reach Planck mass $m_P$. The experimental
challenge would then be to invoke $m_s(T)$ rather than $m_c$.
For the simplified case of
$\beta=0$, 
purely on dimensional grounds, for $m_s \sim m_P$ the departures from 
`$\pm \,(1/2)\, \hbar \omega$' are expected to be of the order 
$m_P\, \omega^2\, \ell_P^2$.\footnote{This estimate is consistent
with calculations given in \cite{VilelaMendes:1999xv}.}
Therefore, by coupling two SQUIDs, set
to \textit{oscillating} super-currents at slightly different
(angular)frequencies $\omega_1$ and $\omega_2$, one may be able to
observe the phase difference
\beq
\frac{1}{\hbar} m_P  (\omega^2_2 -\omega^2_1) \ell_P^2 t\,,
\eeq
thus giving a laboratory signature of quantum gravity.

The cosmological constant, as evaluated with modified zero-point energies
within the context of the proposed Lie algebra for \texttt{IGQR}, 
is likely to remain at variance with data. 
Should this happen, for reasons given in the main text, 
it should be interpreted as 
indicative of the need to replace the suggested Lie algebra
for \textit{IGQR} by its  
supersymmetric version.

\section{Concluding remarks}
\label{Sec:ConcludingRemarks}

The notion of free fall was originally established within 
Newtonian and Galilean framework, and only later did it 
take its new form in  general relativity
by accepting Poincar\'e spacetime symmetries
as symmetries of a freely falling frame. 
For historical reasons, foundational considerations on the subject
which shaped its eventual evolution
were confined to essentially non-quantum realm. 
All problems that came to arise when one confronted 
the interface of gravitational and quantum realms
were, for a long time (with exception of recent years)
considered as technical in nature and were expected to go away. 

A hint that the notion of freely falling frame may be in need
of revision can be deciphered from the following 
observations (which are to be taken in heuristic spirit,
and should not be used against the concrete suggestions
made in section \ref{Sec:FreeFall}):

\begin{enumerate}

\item
There exist freely falling frames
where `gravitationally-induced force'  vanishes while
gravitationally-induced quantum mechanical phases 
do not.
Such considerations can be easily dismissed as the consequent
red-shift of flavour oscillations clocks has no \textit{local}
observability. In challenge, one can argue that
nothing in physics forces one to local observability alone; and that 
this challenge becomes stronger if one considers astrophysical and 
cosmological scenarios. That is, from an operational point
of view freely falling frames exist in which gravitation
cannot be gauged away, though it can be made unobservable for
a local observer\footnote{For a further 
discussion of some these matters, see, e.g.,
\cite{Ahluwalia:1998jx,Konno:1998kq}, and references therein.}.

\item
An \textit{unremovable and intrinsic}
zero-point energy in
freely falling frames also speaks of inherent quantum and gravitational
nature of such frames and that any theory in the \texttt{IGQR}
must respect this circumstance.
The  `120 orders of magnitude' problem associated
with the induced cosmological constant in the standard Heisenberg-Poincar\'e
based framework is perhaps an artifact of the wrong choice 
of algebra associated with spacetime symmetries in
freely falling frames. 

\item
Every consideration on spacetime measurement that allows
gravitational effects asks for non-commutative spacetime
or a modification of spacetime at the Planck scale; see, e.g.,
\cite{Ahluwalia:1993dd,Garay:1994en,Doplicher:1994zv,Sasakura:2000vc,Collins:2004bp}. 
\item
The position and momentum operators commute in the Poincar\'e
algebra, while some of them  do not in Heisenberg algebra.
This \textit{a priori} independence of these two operators
in the Poincar\'e and Heisenberg algebras
in fact already suggests an element of conceptual incompatibility if
one envisages  a unification of these notions. The
suggested 
Lie algebra for \texttt{IGQR} in fact achieves this merging naturally.
Heisenberg and Poincar\'e algebras no longer maintain their 
separate existence but are unified in one stable Lie algebra.
\end{enumerate}

By modifying the notion of a freely falling frame
as carrying a stabilized Heisenberg-Poincar\'e algebra, i.e.,
the Lie algebra for \texttt{IGQR}, to define spacetime symmetries
one generalizes the notion of spacetime where its
characterization requires not only $c$, but also $\hbar$, $\ell_P$,
and $\ell_C$; and possibly $\beta$. 
The small-scale, as defined  by the Planck length $\ell_P$,
and the large scale as encoded in $\ell_C$, are no longer part of
separate realms but intermingle 
in the new notion of 
spacetime symmetries. The resulting spacetime has unavoidable
quantum and gravitational features in which the notion of particles
suffers a modification. Such primitive concepts
as mass and spin no longer remain the same  but 
undergo a well-defined change via Casimir invariants associated with Lie
algebra for \texttt{IGQR}. By modifying the notion of invariant mass,
one's understanding of inertia and equivalence principle shall
require an \textit{ab intio} investigation; while by modifying the
Heisenberg algebra one's notion of quantum undergoes a well-defined,
though not yet fully explored, change.

 \ack
It is my pleasure to thank the following colleagues for
remarks on penultimate draft of
this note: Giovanni Amelino-Camelia, Chryssomalis Chryssomalakos,
Daniel Grumiller, Jerzy Kowalski-Glikman, 
Jerzy Lukierski, Matt Visser, and the three CQG referees for their advice
and questions. 

In part, this work is supported by Consejo Nacional de Ciencia y
Tecnolog\'ia (CONACyT, Mexico) grant E-32067.

\end{document}